\documentstyle[seceq,supplement]{ptptex}

\title{
Gravitational Waves from Relativistic  Stars 
}

\author{
Yasufumi {\sc Kojima}\footnote{
E-mail address: kojima@theo.phys.sci.hiroshima-u.ac.jp}
}

\inst{
Department of Physics, Hiroshima University, 
Higashi-Hiroshima 739-8526
}

\recdate{
\today
}

\abst{%
 Stellar pulsations in rotating relativistic stars 
are reviewed. Slow rotation approximation is applied to solving 
the Einstein equations. The rotational effects on the 
non-axisymmetric oscillations  are explicitly shown in the 
polar and axial modes.
}

\begin{document}
\maketitle

\section{Introduction}

   The amplitude of gravitational waves is quite tiny, 
although they carry enormous amount of energy.
They propagate without suffering any disturbance after 
leaving the sources. The observation gives the direct information 
of the emission regions. The gravitational wave astronomy in the 
next century will therefore be an invaluable tool to diagnose the
interior structure of the relativistic stars.
As one of the promising sources, stellar pulsations are important. 
The oscillations are excited in the newly born neutron stars just 
after supernova explosions, or in abrupt stellar quakes. 
The excitation mechanism is uncertain at present, much more
the expected number of the sources.
The direct observation only will give a clue to these questions.
In addition to the dynamical oscillations, non-axisymmetric 
oscillations play an important role in rotating stars. They cause
secular gravitational radiation-reaction instability,
i.e., CFS-instability \cite{ch70}\tocite{fr78}.
The instability was so far studied in the polar f-modes.
See e.g., Ref.~\citen{li95}. 
Recently, the same mechanism was examined in the axial 
r-modes \cite{an98,frmo98}.
The estimates based on the Newtonian calculations
suggest that the instability sets in  even for 
smaller angular velocity.
The further refinement of the axial modes is therefore crucial.
The r-modes have been examined from various 
aspects \cite{ko98}\tocite{lofr99}. 
The whole of the works can not be reviewed here.
See the some papers and references therein for
the details.

 In this paper, we consider the theoretical aspect of the 
stellar pulsations within general relativity.
Astrophysical applications of the CFS instability 
may be given elsewhere\footnote{
For a recent overview see the paper by Friedman in this volume}.
In \S 2, the mode classification of the stellar pulsations  
is given from the viewpoint of the physical forces 
and mathematical displacements.
This is useful for the following discussion.
In \S 3, the rotational effects on the  polar modes are reviewed.
In \S 4, a method is given to solve the axial modes in
slowly rotating stars.
In \S 5, the method is applied to 
the axial oscillations without metric perturbations. 
The approximation is known as 
the Cowling approximation \cite{co80},
and gives good estimates for some pulsation modes.
In \S 6, the method is  also applied to 
axial oscillations with metric perturbations.
Finally, \S 7 devotes to the summary.
We use the geometrical units of $ c=G=1, $ in this paper.

\section{  Mode classification of pulsations   }

\subsection{  Physical forces  }
    Pulsations reveal various involved physics in different modes
of oscillation patterns and the frequencies.
In order to explain the some pulsation modes, we consider
linearized equations of motion in the Newtonian gravity:
\begin{equation}
 \partial _t \delta v _i +   v ^k \nabla _k  \delta v _i +
 \delta v ^k \nabla _k  v _i 
 =
-  \nabla _i \left( \frac{\delta p}{\rho } \right)
- \frac{ \Gamma p }{\rho^2}
 \left( A_i   \delta \rho + \xi ^k  A_k \nabla _i \rho
 \right) 
-\nabla _i \delta \phi ,
\label{nwt.eqn}
\end{equation}
where $ \Gamma $ is adiabatic index, and $ A_i $
is the Schwarzschild discriminant defined by
\begin{equation}
A_i = \frac{ \nabla _i \rho }{\rho} -
      \frac{ \nabla _i p }{\Gamma p} .
\end{equation}
The first term in the right hand side of eq.(\ref{nwt.eqn}) 
represents the pressure force. This driving force causes the p-mode.
The frequency $\sigma _p$ of the p-mode is typically 
given by the inverse of the free fall time, 
\begin{equation}
  \sigma _p \sim \left( \frac{ M }{ R^3} \right) ^{1/2} ,
\end{equation}
where $ M $ and $ R $ are the mass and the radius of the star.
The frequency increases with the number of nodes.
The f-mode can be regarded as the lowest node-less one.  
The second term in the right hand side of eq.(\ref{nwt.eqn})
is related with the buoyant force. The mode associated with 
the buoyancy is called g-mode. 
The typical frequency $\sigma _g$ of the g-mode is characterized by 
the Brunt-V${\rm \ddot{a}}$is${\rm \ddot{a}}$l${\rm \ddot{a}}$ 
frequency $ \omega _{BV},$ 
\begin{equation}
 \sigma _g \sim \omega _{BV}  = \left(  -A \cdot g  \right) ^{1/2} ,
\end{equation}
where $ g  $ is the local gravitational acceleration.
The value $ A $ significantly depends on the position.  
The convectively unstable region corresponds to
$ A \cdot g  > 0 $ and  stratified region corresponds to
$ A \cdot g  < 0 .$ For isentropic stellar model, 
we have $ A \cdot g  =0 $ and hence  $\sigma _g =0 .$
Since the thermal force is much weaker than 
the dynamical pressure in general, 
the motion driven by the buoyancy  is therefore slow,
and the frequency of the g-mode is much smaller than 
that of the p-mode, 
$  \sigma _g    \ll  \sigma _p . $

These p-, g-modes exist in rather simple situation.
If another physical ingredient is important, what  happens?
New effect changes the modes in general.
For example, stellar rotation changes the frequency as
\begin{equation}
  \sigma = \sigma _0  + \sigma ' \Omega  +  \cdots,
\label{sft.eqn}
\end{equation}
where $ \Omega $ is angular velocity, and
$ \sigma _0 $ is the frequency of the non-rotating star.
In eq.(\ref{sft.eqn}), we  assumed the slow rotation and
considered the first-order effect only. 
In addition to the shift of the frequency, 
the rotation causes a new driving force, i.e., 
Coriolis force. The new mode associated with the force 
is called r-mode. The frequency is given by 
\begin{equation}
    \sigma _r  \sim \Omega .
\end{equation}
In the limit of no rotation, the frequency is zero. 
In this way, a new physical effect causes 
a new mode associated with it.
Another good example of the existence of the new mode 
is w-mode \cite{wmode1}\tocite{wmode4}, which is related 
with the gravitational wave in general relativity.

\subsection{  Oscillation pattern  }
  From the mathematical point of view, 
the oscillation modes can be divided into two classes by parity.
The displacements on a sphere can be expressed by
the 'gradient' or  'rotation' of  the spherical harmonics 
$ Y _{lm} .$ 
The polar (or poloidal) mode  in the spherical star
can be expressed by
\begin{equation}
 \xi_r = R(t,r) Y_{lm}(\theta,\phi),~~~~
 \vec{\xi } = V (t,r)\vec{\nabla} Y_{lm}(\theta,\phi).
\end{equation}
The density and pressure perturbations are coupled in this mode.
Another type of the displacements can be expressed by 
\begin{equation}
  \xi _r =0, ~~~~~
 \vec{\xi } = U (t,r) 
(\hat{r} \times \vec{\nabla} Y_{lm}(\theta,\phi)). 
\end{equation}
This class of displacements is called as the axial (or  toroidal)
mode. The density and pressure perturbations are zero on the 
spherical stars.  Rotation however slightly induces them.

These types of the displacements cause
different types of the gravitational radiation.
The change of the mass moment plays an important role in the 
polar modes, whereas that of the mass current
plays a  role in the axial modes.
Both modes are subject to the CFS instability due to
the radiation reaction.
The polar f-modes work only in the rotating stars with nearly 
break-up speed. On the other hand, recently discovered r-modes 
work even in slow rotation.
The axial mode therefore deserves to be examined more extensively.

%
\section{  Rotational  effects  on  the  polar modes }

In this section, the rotational effects on some p-modes 
are demonstrated. The formulation and the numerical results for 
normal frequencies of the slowly rotating stars are given in 
Refs.~\citen{polar1}-~\citen{polar4}.
The non-radial oscillations with index 
$ l \ge 2 $ are
described by the coupled wave-equations inside the star.
They are schematically written as 
\begin{equation}
( - \partial _t ^2 + \partial _r ^2 ) X + 
F(\partial _r X, \partial _r Y, X, Y) =0,
\end{equation}
\begin{equation}
( - c_s ^2 \partial _t ^2 + \partial _r ^2 ) Y +
 G(\partial _r X, \partial _r Y, X, Y) =0,
\end{equation}
where $ c_s $ is the sound velocity. These two equations
show that the perturbations propagate with 
the light velocity and sound velocity, respectively. 
The set of equations is solved as an eigen-value
problem with 
 $ \exp \{ - i(\sigma t - m \phi) \}. $ 
We impose the out-going wave condition at infinity.
The resultant eigen-value is a  complex number. 
The imaginary part  $\sigma _I $ represents
the decay of the oscillations 
due to the gravitational radiation, if $\sigma _I > 0.$
For the slowly rotating star,
the  frequency can be expanded with the
rotational parameter $ \varepsilon = \Omega \sqrt{R^3/M}. $
As a result, the frequency is modified as
\begin{equation}
\sigma  =\sigma_R ( 1+ m  \sigma_R ' \varepsilon )
 - i \sigma _I ( 1+ m  \sigma_I ' \varepsilon ).
\end{equation}
Note that the axisymmetric mode $(m=0)$ is
affected only from the second-order of the rotation.
That is, the effect of the centrifugal force etc.

Some frequencies of $l=2$ mode are tabulated in Table \ref{tb1}.
The polytropic stellar model is adopted. 
The direct numerical results show that the corrections 
$  \sigma_R ' $ and $   \sigma_I ' $   are positive.
This result means the counter-rotating mode 
$ m < 0 $  beyond the critical velocity changes the sign 
in the pattern speed, i.e., the real part of the frequency 
and in the decay rate,  i.e., the real part of it.
This fact is the condition of the radiation reaction 
instability.  In this way, the counter-rotating mode  
becomes unstable for 
large angular velocity $ \varepsilon . $
Note that the f-mode is the most crucial among the p-modes, 
since the corrections are the largest and the f-mode
becomes unstable first.
The frequencies and the rotational corrections are
calculated for a wide range of stellar models.
The numerical calculations \cite{polar2} show 
that these corrections increase with
the relativistic factor $M/R.$
That is, fully relativistic calculation suggests
that the critical angular velocity decreases.
This means that the instability sets in even for
smaller angular velocity, as the system becomes more 
relativistic.

\begin{table}
\caption[Table]{
Normal frequencies of stellar model with 
$ n=1, M/R =0.2.$ }
\label{tb1}
\begin{center}
\begin{tabular}{ l | c c c c }
\hline \hline
 mode &
 $\sigma_{R} \sqrt{R^3/M}$ &  $ \sigma_{R} '$ &
 $\sigma_{I} R^4/M^3 $ &  $\sigma_{I} '$  \\
\hline
  f     & 1.17 & 0.57 &  0.032 & 3.10 \\
  p$_1$ & 2.70 & 0.32 &  0.006 & 1.78\\
  p$_2$ & 4.12 & 0.12 &  0.001 & 0.90\\
\hline
\end{tabular}\end{center}
\end{table}

\section{   Perturbation scheme  }

   The axial oscillation is trivial in the non-rotating stars,
since there is no restoring force in the fluid stars.
The oscillation becomes possible in the rotating stars.
We therefore have to consider the oscillation on the rotating 
stars, which will be a difficult task.
We will consider the slow rotation approximation in the pulsation
equations.  That is, pulsation equations are expanded with respect to
the rotation parameter. 
The method can be applied to the polar modes successfully.
In the spherically symmetric case, 
the perturbations can be decoupled into 
the axial and polar perturbations with spherical
harmonic index $ (l,m).$  
They are respectively described by 
the axial functions  ${\cal A}_{lm}$ 
$ \equiv  (U_{lm},h_{0~lm},h_{1~lm}),$
and
the polar functions  ${\cal P}_{lm}  $
$ \equiv ( \delta p_{lm}, 
 \delta \rho_{lm}, R_{lm},  V_{lm}, 
 H_{0~lm}, H_{1~lm}, H_{2~lm}, K_{lm}).$ 
In the presence of rotation, the perturbations 
are described by the mixed state of them.   
If the perturbation equations are expanded 
by the rotation parameter $ \varepsilon $, then
the formal relation between the axial-led ${\cal A}_{lm}$
and the polar-led ${\cal P}_{lm}$ can be expressed as 
%
%
\begin{eqnarray}
 0 =  [ {\cal A}_{lm} ] +  
  {\cal E} \times  [ {\cal P}_{l\pm 1m} ]+
  {\cal E}^2 \times [ {\cal A}_{lm} ,  {\cal A} _{l\pm 2m} ]
  + \cdots,
\label{cpl1.eqn}
\\
%
 0 =  [ {\cal P}_{lm} ]+ 
  {\cal E} \times [ {\cal A}_{l\pm 1m} ]+
  {\cal E}^2 \times [ {\cal P}_{lm} ,  {\cal P} _{l\pm 2m} ]
  + \cdots ,
\label{cpl2.eqn}
\end{eqnarray}
where the symbol ${\cal E} $ denotes some functions of order 
$ \varepsilon ,$  and
the square bracket formally represents the relation 
among perturbation functions therein. 
We assume  that  the axial-led and polar-led functions 
are expanded as
\begin{equation}
    {\cal A}_{lm} =  {\cal A}_{lm} ^{(1)}+  
 \varepsilon ^2 {\cal A}_{lm} ^{(2)}  + \cdots,
~~~
    {\cal P}_{lm} =  \varepsilon ( 
{\cal P}_{lm} ^{(1)}+  \varepsilon ^2 {\cal P}_{lm} ^{(2)} +
 \cdots ) .
\end{equation}

Substituting these functions into 
eqs.(\ref{cpl1.eqn})-(\ref{cpl2.eqn}),
and comparing each order of $\varepsilon $,
we have the following equations of $ \varepsilon^n (n=0,1,2)$,
\begin{eqnarray}
0 &=&  [ {\cal A}_{lm} ^{(1)} ],
\label{pts.1}
\\
0 &=&  [ \varepsilon {\cal P}_{l\pm1m} ^{(1)}  +
{\cal E} \times  {\cal A}_{lm} ^{(1)} ],
\label{pts.2}
\\
0 &=&   [  \varepsilon ^2 {\cal A}_{lm} ^{(2)} ]    + 
{\cal E} \times [ \varepsilon  {\cal P}_{l\pm 1m} ^{(1)}]
+ {\cal E}^2 \times 
[ {\cal A}_{l m} ^{(1)},  {\cal A}_{l\pm 2m} ^{(1)} ] 
\label{pts.3}
\\
&=& [ \varepsilon ^2 {\cal A}_{lm} ^{(2)} 
 +
 {\cal E}^2 \times  {\cal A}_{l m} ^{(1)}] .
\label{pts.4}
\end{eqnarray}
We have here assumed that the perturbation is described by 
a single spherical harmonic in the lowest order, that is,   
$    {\cal A}_{l' m} ^{(1)}=0, $
for $ l' \neq l  ,$
and used eq.(\ref{pts.2}) in eq.(\ref{pts.3}).
Equation (\ref{pts.1}) represents the axial oscillation at the
lowest order.
Equation (\ref{pts.4}) is the second-order form of it, and
the term $ {\cal E}^2 \times {\cal A}_{l m} ^{(1)} $
can be regarded as the rotational corrections.
  The method to solve the equations is straightforward.
The first-order equations are solved by the axial-led
functions. The polar-led functions are expressed using them.
We have the second-order equations with the corrections 
expressed by the axial-led functions at the lowest-order.
These equations are successively solved in the following sections.
In the actual calculations, we also assume that
the time variation of the oscillation is slow and
proportional to $\Omega, $ i.e.,
$\partial _t \sim \Omega \sim  O(\varepsilon) .$
This is true in the r-mode oscillation, as will be
confirmed soon.
%

\section{Axial oscillations  in the Cowling approximation }

\subsection{First-order solution }
    We will apply the perturbation scheme described
in the previous section 
to the problem of axial oscillations.
In this section, we neglect the gravitational
perturbation, that is, we consider the Cowling approximation.
The details of the calculations are explained in 
Ref.~\citen{koho99}.
The leading order equation of
$ \delta T ^\mu _{\nu;\mu} = 0$
is reduced to 
\begin{equation}
  ( \partial _T - im \chi ) U_{lm} ^{(1)} = 0 ,
\label{cw.1st}
\end{equation}
where 
\begin{equation}
 \chi = \frac{2}{l(l+1)} \varpi 
      = \frac{2}{l(l+1)} (\Omega -\omega) , 
\end{equation}
and $\partial _T $ denotes time derivative 
in the co-rotating frame, 
i.e., $ \partial _T U_{lm} = (\partial_t + im \Omega) U_{lm} .$
It is easy to observe that 
the motion is trivial for the non-rotating case,
i.e., $ \partial_t  U_{lm} =0.$

If one solves eq.(\ref{cw.1st}) by the eigen-value problem,
then the solution is expressed by a delta function.
Instead, we will examine the evolution of the perturbation by
the initial value problem. 
The solution of eq.(\ref{cw.1st}) is 
\begin{equation}
  U_{lm} ^{(1)} (t,r) = \int  
f_{lm} ^{(1)} (r)  \frac{ e^{st} }{ s + im (\Omega - \chi) } ds = 
f_{lm} ^{(1)} (r) e^{ -im (\Omega-\chi) t} H(t),
\label{sol.1st}
\end{equation}
where $ H(t) $ is the Heaviside step function. 
We will consider $ t > 0 $ region only, so that 
the function $ H(t) $ may well be omitted from now on.
The function  $f_{lm} ^{(1)}$ describes the initial disturbance 
at  $t=0$ .

The implications of the result (\ref{sol.1st}) are as follows.
In the Newtonian stars, the sinusoidal time-dependence
can be described by a single frequency,
\begin{equation}
 m( \Omega - \chi) \to 
 \sigma _N =\left( 1- \frac{2}{l(l+1)} \right ) m \Omega .
\end{equation}
This is the r-mode frequency measured in the non-rotating frame.
In the relativistic stars, 
$\varpi $  is monotonically increasing 
function of $r $,  $\varpi _0 \leq  \varpi \leq \varpi_R .$ 
The possible frequency range is then spread out,
\begin{equation}
  \left( 1- \frac{2}{l(l+1)} \frac{\varpi _R}{\Omega} 
   \right )m \Omega  
   \leq  \sigma   \leq
   \left( 1- \frac{2}{l(l+1)} \frac{\varpi _0}{\Omega} 
    \right ) m \Omega  .
\end{equation}
In both cases, Newtonian and relativistic stars, 
the radial dependence $ f_{lm} ^{(1)}  $ is arbitrary at this order. 
The function $f_{lm} ^{(1)}  $ in eq.(\ref{sol.1st})  
is constrained by the equation of motions for the polar part,
as will be shown in the subsequent subsections.
In this meaning, the scheme (\ref{pts.1})-(\ref{pts.4})
is degenerate perturbation scheme.
%

\subsection{  Second-order equations  }

According to (\ref{pts.2}),
the polar-led functions 
$ ( \delta p_{l\pm1 m} ,  \delta\rho_{l\pm1 m},
    R _{l\pm1 m},   V _{l\pm1 m} )$
are  expressed by  $  U_{lm} ^{(1)} .$ 
These corrections affect the axial parts with indices 
$ (l\pm2, m) $ and $(l, m).$ 
We consider the equation with $(l, m)$ only. 
Corresponding to eq.(\ref{pts.4}),
the axial equation with the corrections up to 
$O(\varepsilon^3)$ is 
\begin{equation}
 0  = (\partial_T  - im \chi ) U_{lm} ^{(2)} 
  + {\cal L}[ \partial_T U_{lm} ^{(1)} ],
\label{cw.2nd}
\end{equation}
where 
$ {\cal L} $ is the Sturm-Liouville differential operator
defined by
\begin{eqnarray}
 {\cal L} [\partial_T U_{lm} ^{(1)} ]
  &=&
   8 c_3 \varpi e^{(-\lambda/2 - \nu)}
   (\rho_0 + p_0)
   \left[
    \frac{r^2e^{(\lambda - \nu)/2}}{A\nu'(\rho_0 + p_0)}
    (e^{\nu/2} \varpi \partial_T U_{lm} ^{(1)} )'
    \right]'
\nonumber
\\
&&~~~
  - (F + G) \partial_T U_{lm} ^{(1)} ,
\\
F &=&
    -4 c_3 \varpi^2 e^{(-\lambda/2 - \nu)}
    \left(
     \frac{r^2 e^{\lambda/2} \rho_0 '}{A p_0'}
     \right)' 
\nonumber
\\
&&~
  - \frac{4 c_2 \varpi^2 e^{(-\lambda - 3\nu)/2}}{\rho_0 + p_0}
    \left[
    \frac{e^{(\lambda + 3\nu)/2}}{A\nu'\varpi}
    (\varpi r^2 e^{-\nu})'
    (\rho_0 + p_0)
     \right]'
\nonumber
\\
&&~ 
  +   \left(
    \frac{2 c_1 e^{\nu}}{A\nu'r^2}
    \right)
   [(r^2 \varpi e^{-\nu})']^2  ,
\\
G &= &
   -8 c_3 \varpi^2 e^{(-\lambda/2 - \nu)}
    \left(      \frac{r^2 e^{\lambda/2} }{\nu '}
    \right)' 
  + \frac{4 c_2  }{\nu' r^2} (\varpi ^2  r^4 e^{-\nu} )'
\nonumber
\\
&&~
- 3 c_1
 \left[
  r e^{-\lambda /2 } 
  \left( \frac{ e^{\lambda /2 } \xi_2 }{r} \right) '
  -  \frac{3}{2}\frac{\varpi'}{\varpi}\xi_2
  -  k_2   
 + \frac{e^\lambda}{r} m_2
 + \frac{5W_3}{\varpi}
 \right ]
\nonumber
\\
&&~
 - \frac{3 m^2}{l(l+1)}
 \left[
  \frac{\xi_2}{r}
  + \frac{1}{2}
   \frac{\varpi'}{\varpi}
    \xi_2
  +    k_2 - \frac{5W_3}{\varpi}
  \right] 
 -\left(   \frac{W_1}{\varpi}  +  \frac{6W_3}{\varpi}  \right ).
\end{eqnarray}

\subsection{ 
Solution of the radial function  
- mode specification - 
}

In order to solve eq.(\ref{cw.2nd}),
we introduce a complete set of functions 
 $ y_\kappa (r) $ with \\
eigenvalue   $ - \kappa $  with respect to
the Sturm-Liouville operator ${\cal L }$,
\begin{equation}
   {\cal L } [ y_\kappa   ] + \kappa  y_\kappa  =0  .
\label{eqn.st}
\end{equation}
The second-order equation  (\ref{cw.2nd}) can be 
integrated with  
 $y_\kappa =im \chi f_{lm; \kappa} ^{(1)} (r) $
and unknown function  $f_{lm} ^{(2)} (r) $  as 
\begin{equation}  
 U_{lm} ^{(2)} = 
 ( i m \kappa  \chi  t f_{lm; \kappa} ^{(1)} +f_{lm} ^{(2)} )
  e^{-im(\Omega -\chi)t } .
\end{equation}
The sum of the first and second order forms is
approximated as
\begin{eqnarray}
U_{lm} ^{(1)} + U_{lm} ^{(2)}  &=&
\left[ (1 +  i m  \kappa  \chi t ) f_{lm; \kappa} ^{(1)} 
+f_{lm} ^{(2)} \right] e^{ -im (\Omega - \chi) t} 
\label{cw.sum1}
\\
& = &
 \left[ f_{lm; \kappa} ^{(1)} + f_{lm} ^{(2)} \right]
   e^{ -im( \Omega -(1 + \kappa ) \chi ) t } ,
\label{cw.sum2}
\end{eqnarray}
where we have exploited the freedom of $ f_{lm} ^{(2)} $ 
to eliminate the unphysical growing term in eq.(\ref{cw.sum1}). 
The value $ \kappa $ originated from fixing of the initial data
becomes evident for large $t$,
since the accumulation of small effects from the higher order 
terms is no longer neglected.
As a result, the frequency should be adjusted
with the second-order correction
to be a good approximation even for slightly 
large $t$, as  eq.(\ref{cw.sum2}). 
This remedy is known as 
the renormalization of the frequency, or
strained coordinate for $t$ in the perturbation 
method \cite{hi91}.

  We here summarize the calculations up to the second-order
by the perturbation scheme (\ref{pts.1})-(\ref{pts.4}).
In the lowest-order calculation, 
the first-order spatial function is not determined, whereas
the time-dependence is fixed.
By considering the next order, 
the first-order spatial function 
$f_{lm; \kappa} ^{(1)} $ is specified with 
the second-order correction of the frequency $  \kappa .$
The range and the nature of the correction $  \kappa $
can not be explored without explicitly solving 
the eigen-value problem (\ref{eqn.st}), which  
significantly depends on the equilibrium state,
in particular, $ A= 0 $ or $ A \ne 0 $.
For the barotropic case ($ A= 0 $), 
eq.(\ref{cw.2nd}) is modified \cite{koho99}.
%

\section{Axial oscillations including gravitational perturbations}

\subsection{ Lowest-order calculation }
In this section, we incorporate the metric perturbations.
The method to solve the axial oscillations is the same as in 
the Cowling approximation. We have additionally 
six metric functions $(h_{0~lm}, h_{1~lm},
 H_{0~lm},   H_{1~lm},   H_{2~lm}, K_{lm}).$
In the lowest-order, the calculation is rather simple,
since  only two components, $ h_{0~lm} $ and $ h_{1~lm} $ 
are relevant.
We define a function $ \Phi _{lm} $ as 
\begin{equation} 
  \Phi_{lm}  = \frac{h_{0~lm} }{r^2} .
\end{equation}
The relation between the metric functions is given by
\footnote{ There is a misprint in the previous paper
\cite{ko98}. }
\begin{equation}
 h_{1~lm} =  { r^4 e^{-\nu} \over  (l-1)(l+2) }
\left[ ( \partial _T  - i m \varpi ) \Phi '_{lm}
    -{ 2 im \omega ' \over l(l+1)} \Phi _{lm}
\right] .
\end{equation}
The axial velocity function is expressed by two ways: 
\begin{eqnarray}  
  \left( \partial _T  - i m \chi \right) U _{lm}
& = & -4 \pi (\rho +p) r^2 e^{-\nu} \partial _T 
  \Phi _{lm},
\label{mt1.tm}
\\ 
U _{lm} & = & \frac{r^2 j^2 }{4} 
 \left[ \frac{1}{j r^4}
 \left( j r^4    \Phi  _{lm}  ' \right)' 
 - ( v + 16\pi (\rho +p)  e^{\lambda} )    \Phi  _{lm}
\right],
\label{mt1.sp}
\end{eqnarray}
where
\begin{eqnarray}
  v &=&  \frac{e^{ \lambda }}{r^2} \left[
       l(l+1) -2 \right] , 
\\
 j &=& e^{- (\lambda +\nu )/2} .
\end{eqnarray}
Eliminating $U _{lm} $ in eqs.(\ref{mt1.tm})-(\ref{mt1.sp}),
we have the master equation as
\begin{equation}  
  \left( \partial _T  - i m \chi \right)
 \left[ \frac{1}{j r^4}
 \left( j r^4   \Phi_{lm}   ' \right)' 
 - v    \Phi_{lm}  \right]  = 
    - 16 \pi i m  \chi (\rho +p ) e^{ \lambda } 
 \Phi _{lm} .
\end{equation}
This equation is reduced to an eigen-value 
equation \cite{ko98}, by assuming $ \exp(-i\sigma t ):$ 
\begin{equation}  
  \left(  \varpi - \mu \right)
  \left[ \frac{1}{j r^4}\left( j r^4   \Phi _{lm}  ' \right)'
  - v    \Phi _{lm}  \right]   = q \Phi _{lm},
\label{eqn.ray}
\end{equation}
where 
\begin{eqnarray} 
  \mu &=&
 - \frac{ l(l+1) }{2m}(  \sigma -m \Omega ),
\\
  q & = & \frac{1}{j r^4} \left( j r^4  \varpi ' \right)' 
    = 16 \pi (\rho +p ) e^{ \lambda }  \varpi \geq 0 .
\end{eqnarray}
Equation (\ref{eqn.ray}) is 
called singular eigen-value equation,
since it has a singular point $r_0 $ unless
$q(r_0) = 0,$ corresponding to
the real value of $ \mu =  \varpi (r_0) .$
This kind of singular eigen-value equation is
studied for the incompressible shear flow.
See the Appendix for the Rayleigh equation
in two dimensional parallel flows.
Some important conclusion can be 
derived from the behavior of the background flow,
 i.e., the function $ \varpi $ in this problem.
For example, the necessary condition of the instability is
that the function $\varpi $ has inflection point, i.e.,
$ q =0. $  
This condition is never satisfied inside the star.
As a result, the frequency is real and the flow is stable
in this order.

\subsection{ 
From the Rayleigh  to Orr-Sommerfeld equations
}

In this section, the second-order corrections will be included.
The calculation is straightforward, but very complicated in actual.
We only consider the term with the highest rank of derivative
with respect to $r$. The rank of the derivative
is important factor to determine the type of equations. 
The term is originated from  the second-order
derivative of $ U_{lm} $ as eq.(\ref{cw.2nd}).
Since  $ U_{lm} $ is expressed by the second-order derivative
of $ \Phi_{lm} $, we have the fourth-order derivative of
$ \Phi_{lm} .$
The term with the highest rank of derivative is explicitly 
given by
\begin{equation}
 {\cal D}_o  [\Phi_{lm} ^{(1)}]  = 8 c_3 \frac{ \varpi e^{ -\nu} }{j r^2}
    \left \{
     \frac{ (\rho_0 + p_0) r^2}{ j A \nu '}
       \left[
         \frac{ j \varpi }{ (\rho_0 + p_0) r^2}
          \left( j r^4 \partial _T \Phi _{lm} ^{(1) \prime} \right)'
        \right]' 
     \right \}' .
\label{df4th}
\end{equation} 
The second-order equation with  this correction 
can be written as 
\begin{equation}  
  \left( \partial _T  - i m \chi \right)
 \left[ \frac{1}{j r^4}
 \left( j r^4   \Phi _{lm}  ^{(2) \prime} \right)' 
 - v    \Phi _{lm} ^{(2)}  \right]  = 
    - 16 \pi i m  \chi (\rho +p ) e^{ \lambda } 
 \Phi _{lm} ^{(2)} + {\cal D} _o [ \Phi _{lm} ^{(1)} ] .
\label{grv2nd}
\end{equation}  
The term  (\ref{df4th}) effectively gives the 'viscosity' like 
the Orr-Sommerfeld equation in the incompressible shear flow.
(See Appendix.)
The viscosity is important for the stability of the flows.
For the small Reynolds number, the laminar flow is realized,
whereas the flow becomes  turbulence above 
a critical Reynolds number.
The effective Reynolds number $ R_e $ in eq.(\ref{grv2nd})
is estimated from dimensional argument as 
\begin{equation}
  R_e \sim \frac{A \nu '}{ \varpi ^2} \sim 
 \frac{ \omega_{BV} ^2 }{ \varpi ^2}  .
\end{equation} 
The viscosity term will play
a key role on the singular point of the first-order
equation, but the consequence is not clear at moment.
It is necessary to explore further how the effective 
Reynolds number should operate in the stability and so on.
%

\section{ Summary
- pulsations and gravitational waves -
}

The polar modes, which exist in the spherical non-rotating stars,
are extensively studied so far. 
The angular dependence of the modes is specified by a single 
spherical harmonic index. The decoupled radial equation can be 
calculated as an eigen-value problem.
The rotational effects are also examined within the first-order,
and the corrections are calculated as
$ \sigma = \sigma _0  + m \varepsilon^1 \sigma _1 + \cdots . $
The relation between the pulsation and the
gravitational radiation is evident in the modes,
that is, the gravitational emission gives the imaginary part 
of the frequency.
The gravitational waves would therefore
give a good insight into the stellar interiors, when observed.

Axial modes never exist in the non-rotating fluid stars, but
exist in rotating stars. The oscillations are calculated for
the slow-rotation approximation. 
When  the mode is calculated by the eigen-value problem, 
large number of spherical harmonics are required in general 
\cite{lofr99}.
In this paper, we considered the mode whose angular 
dependence is dominated for a single spherical harmonic.
As a result,  
the solution is constructed by an initial-value problem.
The temporal dependence can be written as 
an infinite sum of the Fourier mode $ \exp( -i \sigma t ) .$
The rotational corrections are calculated up to the 
third-order as
$ 
\sigma = \varepsilon^1 \sigma _1 (r)  +
 \varepsilon^3  \sigma _3 (r) + \cdots .
$
At present, the relation to the gravitational radiation
is not clear. From the Newtonian estimates \cite{frmo98}, 
the radiation reaction affects the oscillations in order 
$ \varepsilon ^{2m+2}.$
The reaction term should be even power of $ \varepsilon, $
which is time-asymmetric and originated from
the radiative boundary condition at infinity. 
Rather, the equations describing the axial
oscillations are clearly related with the vortex as
the Rayleigh or Orr-Sommerfeld equations.
%

\section*{Acknowledgements}
This was supported in part 
by the Grant-in-Aid for Scientific Research Fund of
the Ministry of Education, Science and Culture of Japan
(08640378, 08NP0801).

\appendix
\section{ Incompressible shear flow}

In this appendix, linear perturbation equations
of the two-dimensional flows are summarized 
for the comparison.
The detailed arguments can be written in some
books \cite{drre81} of hydrodynamics. 
The basic parallel flow of incompressible matter
is assumed to $(0, u(x)).$
We consider the perturbation within the inviscid theory,
and assume the stream function in the form as
$ \Phi (x) \exp \{ - ik (y- ct) \}, $ then we have 
\begin{equation}  
 \left(  u-c \right) \left[ \Phi '' -k^2 \Phi \right] 
 = u'' \Phi .
\label{ray1}
\end{equation}
This is called Rayleigh's stability equation.
Some theorems are proved for the equation. For example,
{\it Rayleigh's inflexion-point theorem } is 
stated as follows.
A necessary  condition for instability is that the 
basic velocity profile should have an inflexion point, i.e.,
$u''=0 .$
However this condition is not sufficient.
The sufficient condition is not yet found.

Viscosity can not be neglected in some flows. 
If we consider the problem within the viscous theory,
the perturbation equation of the two-dimensional  flows
is reduced to the Orr-Sommerfeld equation. 
Using  appropriate  normalization, we can express it as
\begin{equation}  
 \left( \partial _t + u \partial _y \right)
 (\partial _x ^2 + \partial _y ^2)  \Phi =  u'' \partial _y \Phi  
+ R_e ^{-1} (\partial _x ^2 + \partial _y ^2)^2 \Phi .
\label{orsom1}
\end{equation}
In eq.(\ref{orsom1}), $ R_e  $ is the Reynolds number 
defined by $ R_e = u_0  \ell / \nu ,  $
where  $ \nu $ is kinematical viscosity, and 
$ u_0  $ and  $ \ell $ are typical scales of the velocity 
and length. 
Using the Fourier mode, we have
\begin{equation}  
 \left(  u-c \right) \left[ \Phi '' -k^2 \Phi \right] 
 = u'' \Phi  - \frac{1}{i k R_e}
\left[ \frac{d^2}{dx^2} -k^2  \right] ^2 \Phi .
\end{equation}
Formally, the Rayleigh equation (\ref{ray1}) can be regarded as
the leading term in the expansion of $ ( k R_e )^{-1} $.
The Rayleigh equation is a good approximation, 
if the second and fourth order derivatives can be neglected.
When the condition is no longer valid in some regions, 
we have to take account of the viscosity term there.
The stability of the viscous flow significantly depends on
the Reynolds number. For the flow with low 
Reynolds number, the flow is laminar, whereas
it becomes turbulent for the high Reynolds number.
More detailed arguments with boundary conditions
are indispensable to find out 
the critical number, flow pattern and so on.

\end{document}